\documentclass[%
superscriptaddress,
preprint,
nofootinbib,
 amsmath,amssymb,
 aps,
prb,
]{revtex4-1}
\usepackage[section]{placeins}
\usepackage{graphicx}
\usepackage{dcolumn}
\usepackage{bm}

\usepackage[version=3]{mhchem}
\usepackage{color}
\usepackage{float}
\usepackage{bigstrut}

\begin{document}
\sloppy

\title{Superconductivity in rubidium substituted \ce{Ba_{1-\textit{x}}Rb_{\textit{x}}Ti2Sb2O}}

\author{Fabian von Rohr}
\email{vonrohr@physik.uzh.ch}
\affiliation{Physik-Institut der Universit\"at Z\"urich, Winterthurerstrasse 190, CH-8057 Z\"urich, Switzerland}
\affiliation{Laboratory of Inorganic Chemistry, ETH Z\"urich, Wolfgang-Pauli-Str. 10, CH-8093 Z\"urich, Switzerland}

\author{Reinhard Nesper}
\affiliation{Laboratory of Inorganic Chemistry, ETH Z\"urich, Wolfgang-Pauli-Str. 10, CH-8093 Z\"urich, Switzerland}

\author{Andreas Schilling}
\affiliation{Physik-Institut der Universit\"at Z\"urich, Winterthurerstrasse 190, CH-8057 Z\"urich, Switzerland}

\date{\today}

\begin{abstract} We report on the synthesis and the physical properties of \ce{Ba_{1-\textit{x}}Rb_{\textit{x}}Ti2Sb2O} ($x \leq 0.4$) by x-ray diffraction, SQUID magnetometry, resistivity and specific heat measurements. Upon hole doping by substituting Ba with Rb, we find superconductivity with a maximum $T_{\mathrm{c}} =$ 5.4 K. Simultaneously, the charge-density-wave (CDW) transition temperature is strongly reduced from $T_{CDW} \approx$ 55 K in the parent compound \ce{BaTi2Sb2O} and seems to be suppressed for $x \geq 0.2$. The bulk character of the superconducting state for the optimally doped sample ($x$ = 0.2) is confirmed by the occurrence of a well developed discontinuity in the specific heat at $T_{\mathrm{c}}$, with $\Delta C/T_{\mathrm{c}} \approx $ 22 mJ/mol K$^2$, as well as a large Meissner-shielding fraction of $\approx$ 40 \%. The lower and the upper critical fields of the optimally doped sample ($x$ = 0.2) are estimated to $\mu_0 H_{c,1}(0) \approx$ 3.8 mT and $\mu_0 H_{c,2}(0) \approx$ 2.3 T, respectively, indicating that these compounds are strongly type-II superconductors.
\end{abstract} 

\maketitle


\section{Introduction}
Density waves (DWs) are collective states of broken symmetry that arise from electronic instabilities that are often present in highly anisotropic structures. In some cases they compete with superconductivity, which is an another collective electronic state. The emergence of superconductivity in iron arsenides has attracted great interest in the physical properties of transition metal pnictides in general. In most of these materials, superconductivity occurs in proximity to a spin-density-wave (SDW) transition, and is found to exhibit unconventional properties (see, e.g., references \onlinecite{Zhao} and \onlinecite{CaFe2As2}). Until now, several systems have been investigated in which superconductivity emerges upon complete or partial suppression of charge-density-wave (CDW) ordering, such as 2H-NbSe$_2$ \cite{NbSe2}, \ce{Ba_{1-\textit{x}}K_{\textit{x}}BiO3} \cite{Cox,Hamann}, and \ce{Cu_xTiSe2} \cite{CuxTiSe2}. Likewise, the formation of CDW order has been observed in the normal state of superconducting \ce{YBa2Cu3O_{6.67}}.\cite{YBCO_DW} \\
Most recently, \ce{BaTi2Sb2O} has been found to intrinsically host both distinct states of broken symmetry, showing a CDW ordering transition at $T_{CDW} \approx$ 55 K, as well as a transition to superconductivity at $T_c \approx$ 1 K.\cite{BaTi2Sb2O} \ce{BaTi2Sb2O} belongs to a large family of stacked, layered titanium oxide pnictide compounds. In these materials the nominal valence of titanium is Ti$^{3+}$ with its 3d orbitals singly occupied. In this 3d$^1$ configuration, Ti is surrounded octahedrally by O and Sb, forming a \ce{Ti2O} square sublattice. From a structural and chemical perspective, these \ce{Ti2O} layers can be interpreted as the 3d$^1$-anticonfiguration of the 3d$^9$ \ce{CuO2} planes in the cuprates. \\
Upon substitution of barium by sodium or potassium, or of antimony by bismuth or tin, in \ce{BaTi2Sb2O}, the CDW ordering temperature $T_{CDW}$ in substituted \ce{BaTi2Sb2O} is lowered, while superconductivity reaches a maximum $T_c$ of up to 5.5 K at substitution levels around 15 mol-\%. \cite{Doan_JACS,Johrendt_K,bismuth,tin} Detailed NMR and $\mu$SR studies have shown that a CDW ordering is most likely competing and coexisting with a conventional superconductor with an $s$-wave gap. \cite{NMR,fvrohr} \\
In this article we will show how superconductivity and CDW ordering evolve in \ce{Ba_{1-\textit{x}}Rb_{\textit{x}}Ti2Sb2O} as a function of the substitution of barium by rubidium. We will evidence that the CDW ordering transition $T_{CDW}$ is continuously lowered and eventually suppressed with increasing rubidium content, while the transition temperature to superconductivity, $T_c$, is increased, reaching a maximum $T_{c,max}$ = 5.4 K for $x$ = 0.2 (in specific-heat measurements). Our results support the idea that the chemical pressure effect for the superconductivity in these materials is of little importance, whereas the hole doping by the incorporation of Na$^+$, K$^+$, and Rb$^+$, and along with it the variation of the density of states at the Fermi-level $D(E_\mathrm{F})$, appears to be of great importance for the occurrence of CDW ordering and/or superconductivity. Our results further support that \ce{BaTi2Sb2O} is a versatile model-system for the investigation of the competition and coexistence of conventional superconductivity and CDW ordering.

\section{Experimental}
Solid-state reactions were employed to synthesize polycrystalline samples of \ce{Ba_{1-\textit{x}}Rb_{\textit{x}}Ti2Sb2O} with $x$ = 0, 0.02, 0.05, 0.1, 0.15, 0.2, 0.25, 0.3 and 0.4. BaO (99.99\%), \ce{Rb2O} (95\%), Ti (99.99\%), \ce{TiO2} ($>$ 99\%) and Sb (99.999\%) were mixed according to the stoichiometric ratio, ground thoroughly, and pressed into pellets of approximately 1 g batches in an argon filled glove box. The pellets were sealed in argon filled (1 atm) niobium ampules and then sintered at 1000 \nolinebreak[4] $^{\circ}$C for 72 h. Then the samples were reground under inert atmosphere, repelletized and sintered again for 24 h at 1000 \nolinebreak[4] $^{\circ}$C.\\
\begin{figure} [t]
\centering
\includegraphics[width=0.7\linewidth]{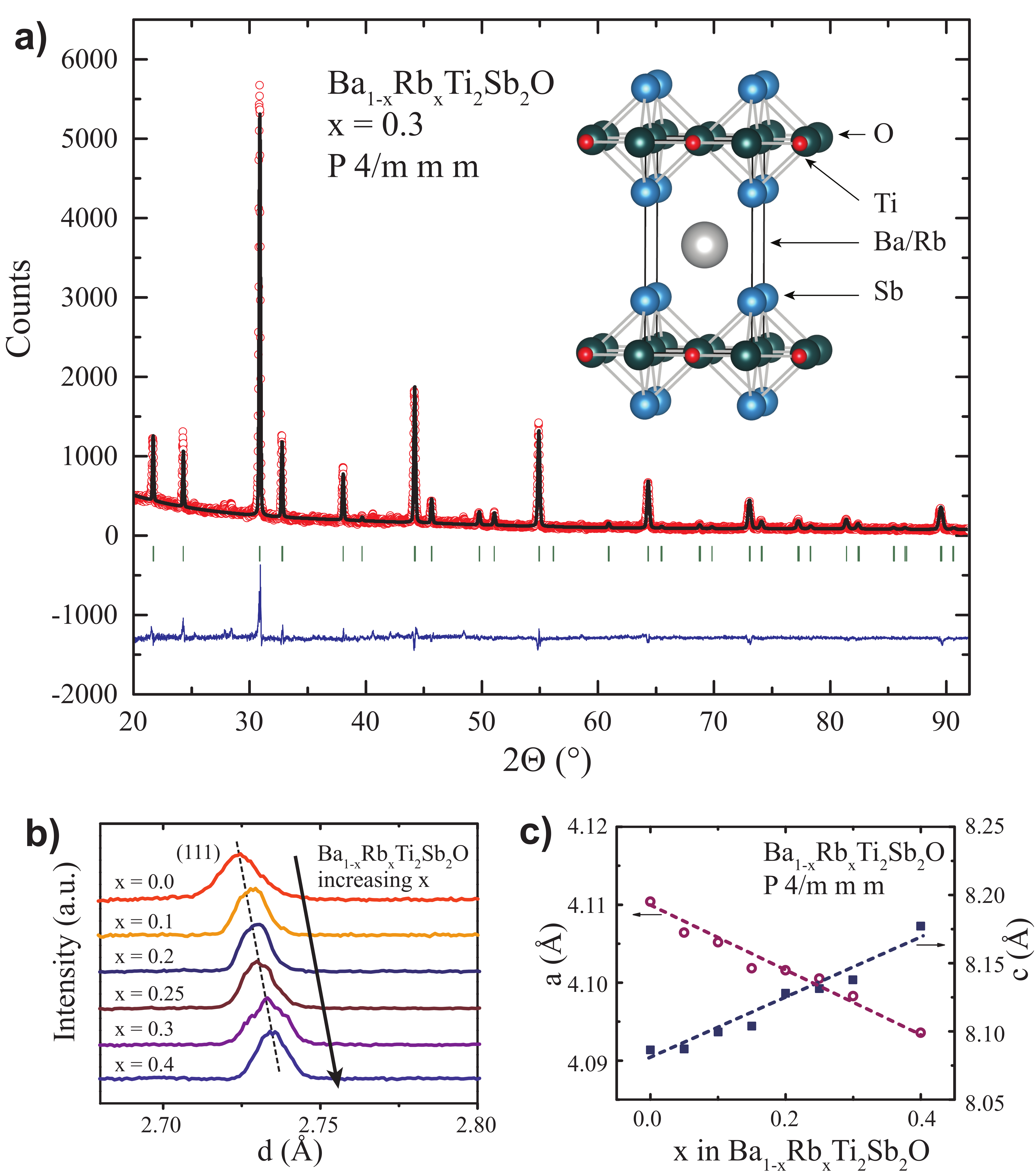}
\caption{(a) The powder x-ray diffraction pattern at ambient temperature for the sample of nominal composition \ce{Ba_{0.7}Rb_{0.3}Ti2Sb2O} ($x$ = 0.3). The crystal structure (\ce{CeCr2Si2C}-type) of the compound is shown in the inset. The vertical dark green lines show the theoretical Bragg peak positions for this phase. The blue pattern on the bottom is the difference plot between the theoretical pattern and the observed intensities. (b) Enlarged XRD data around the (111) reflections as functions of increasing $x$, demonstrating the change of the lattice parameters with Rb content. The dotted line is a guide to the eye. (c) Cell parameters for the different compositions used in this study. The dashed line represents an idealized linear change of the cell parameters.}
\label{fig:1}
\end{figure}
X-ray powder diffraction measurements were performed using a Stoe STADIP diffractometer (Cu-$\mathrm{K_{\alpha 1}}$ radiation, $\lambda$ = 1.54051 \nolinebreak[4] $\mathrm{\AA}$, Ge-monochromator). Rietveld refinements and profile fits were performed using the FullProf program.\cite{Fullprof} The magnetic properties were studied using a Quantum Design Magnetic Properties Measurements System (MPMS XL) equipped with a reciprocating sample option (RSO). The plate like samples were placed in parallel to the external magnetic field in order to minimize demagnetization effects. Specific heat and resistivity measurements were performed with a Quantum Design Physical Property Measurement System (PPMS). For the resistivity measurements, a standard 4-probe technique was employed with 50 $\mu$m diameter gold wires attached with silver paint. The applied current for these measurements was $I$ = 1.5 mA. Specific-heat measurements were performed with the Quantum Design heat-capacity option using a relaxation technique.
\section{Results and Discussion}
In figure \ref{fig:1}a we show the x-ray diffraction pattern at ambient temperature for the sample of nominal composition \ce{Ba_{0.7}Rb_{0.3}Ti2Sb2O} ($x$ = 0.3), together with the crystal structure of the compound in the inset. All compounds of the \ce{Ba_{1-\textit{x}}Rb_{\textit{x}}Ti2Sb2O} ($x \leq 0.4$) solid solution were found to belong to the same crystal structure isopointal to the \ce{CeCr2Si2C}-type structure (\textit{P}4/\textit{mmm}). All experimentally observed intensities are in good agreement with the theoretical pattern, indicating the validity of the present structural model. The slight systematic variation of the (111) reflection as a function of rubidium content $x$ is displayed in figure \ref{fig:1}b. Within the \ce{Ba_{1-\textit{x}}Rb_{\textit{x}}Ti2Sb2O} solid solution $0 \leq x \leq 0.4 $ the cell parameter vary only slightly, but continuously with increasing rubidium content. This can be seen from the small change of $a$ and $c$ obtained from profile fits as shown in figure \ref{fig:1}c. The apparent deviation from a linear variation of the cell parameters with $x$ can be attributed to small deviations from the precise stoichiometry, since we used here the nominal $x$ values for all samples. The trend of a contraction of the \ce{Ti2O} sheets (decreasing $a$) and a simultaneous elongation of $c$ with increasing $x$ is in line with observations on potassium substituted \ce{BaTi2Sb2O}.\cite{Johrendt_K}\\
\begin{figure}
\centering
\includegraphics[width=0.7\linewidth]{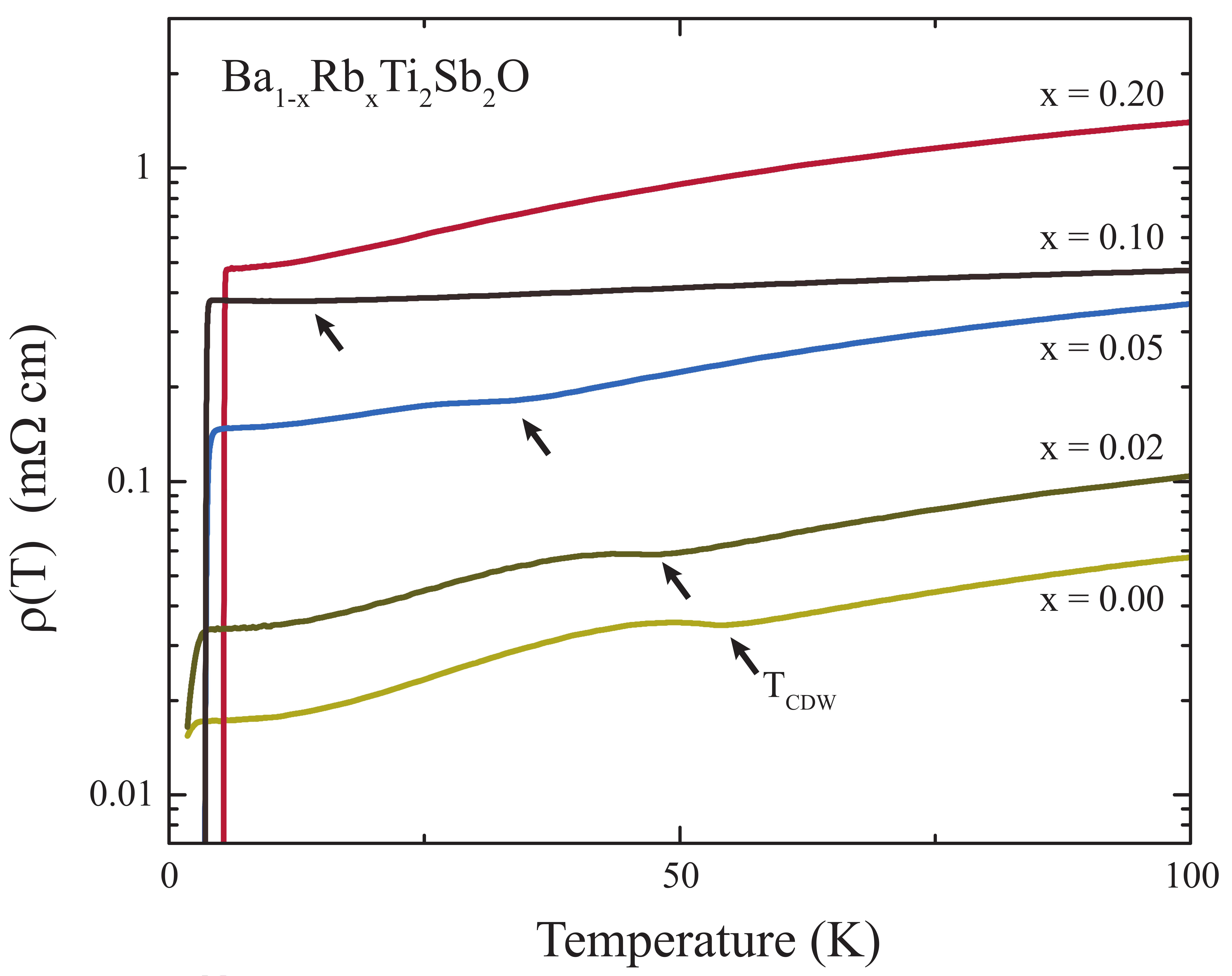}
\caption{Resistivities of \ce{Ba_{1-\textit{x}}Rb_{\textit{x}}Ti2Sb2O} for $x$ = 0, 0.02, 0.05, 0.1, and 0.2 in a temperature range between 1.8 K and 100 K. The CDW transition manifests itself in a sudden increase of the resistivity for the samples x = 0, 0.02, 0.05 and 0.1 (marked with a black arrow). This anomaly is absent for the sample with x = 0.2. For increasing rubidium contents $x$ (up to $x$ = 0.2), the normal state resisitivities increase, the CDW ordering transition temperatures $T_{CDW}$ decrease, while the critical temperatures $T_c$ increase.}
\label{fig:2}
\end{figure}
The resistivities $\rho(T)$ in a temperature range of $T =$ 1.8 K to 100 K are shown in figure \ref{fig:2} for $x$ = 0, 0.05, 0.1, and 0.2. All compounds are metals with resistivities $\rho$(100 K) varying from 0.057 $m\Omega$ cm$^{-1}$ for the parent compound to 0.575 $m\Omega$ cm$^{-1}$ for $x$ = 0.2. The parent compound ($x$ = 0), shows a distinct kink in the resistivity at $T_{CDW}$ = 55 K, which has earlier been attributed to a CDW transition. \cite{fvrohr} This phase transition temperature $T_{CDW}$ is strongly reduced and eventually suppressed with increasing rubidium content. For a relatively small doping of $x$ = 0.05 the CDW transition is already lowered to 34 K, whereas the critical temperature is increased by more than 2 K to $T_c$ = 3.7 K. Taking the absolute value of the resistivity as a measure for the metallicity of a sample, we can state that the increase of the superconducting transition $T_c$ 4 and the decrease and subsequent suppression of the CDW ordering temperature $T_{CDW}$ go along with a decrease in metallicity. A similar trend for $\rho(T)$ has been reported for Na and Bi substituted \ce{BaTi2Sb2O}, and it has been attributed to enhanced impurity scattering or to the proximity to a metal-insulator transition. \cite{bismuth,Gooch13}

\begin{figure}
\centering
\includegraphics[width=0.7\linewidth]{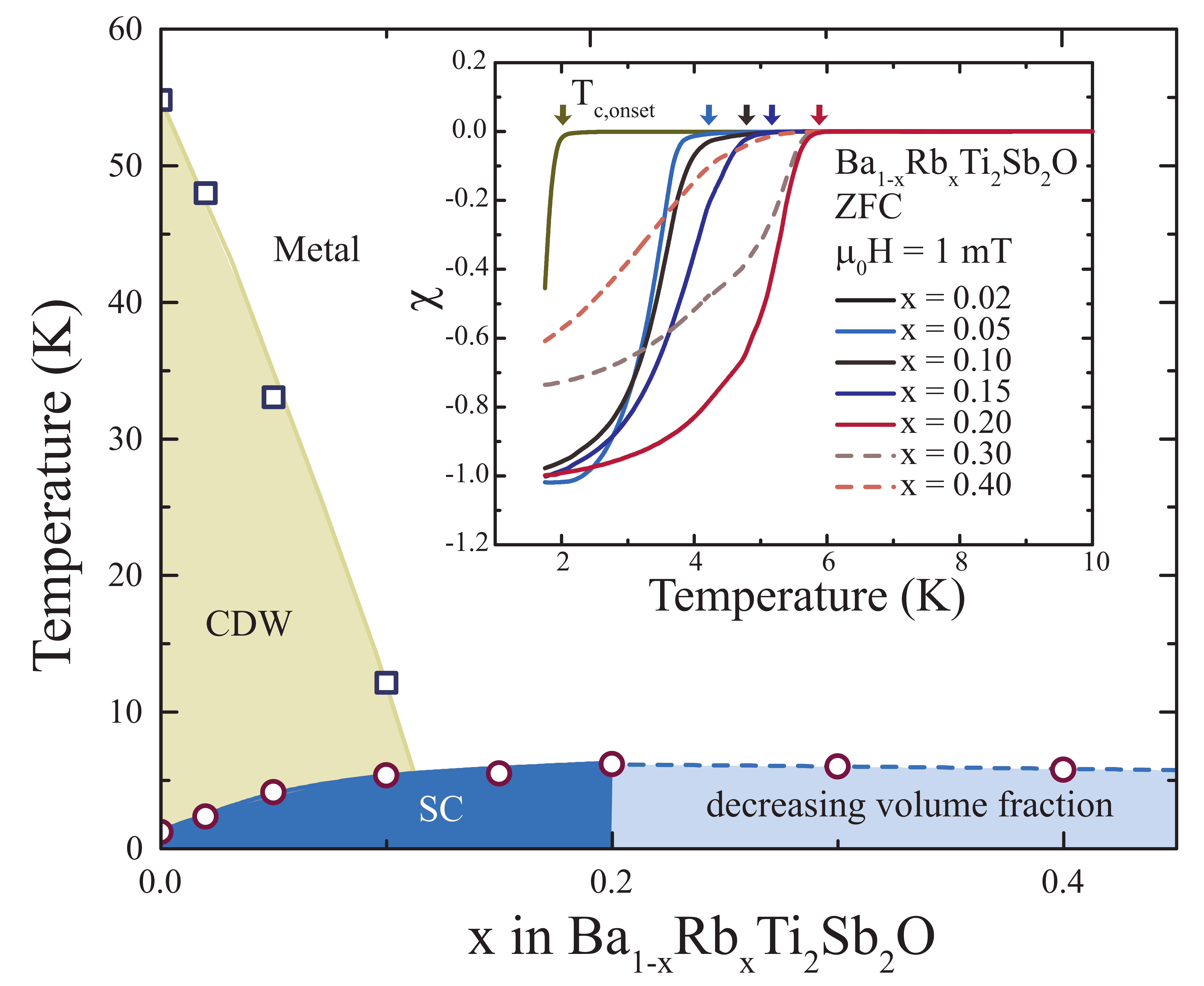}
\caption{Phase diagram of the electronic properties of \ce{Ba_{1-\textit{x}}Rb_{\textit{x}}Ti2Sb2O} derived from resistivity and magnetization measurements ($T_c$ for the parent compound $x$ = 0 was taken from reference \onlinecite{BaTi2Sb2O}). The inset shows the magnetic susceptibility of \ce{Ba_{1-\textit{x}}Rb_{\textit{x}}Ti2Sb2O} for $x$ = 0.02, 0.05, 0.1, 0.15, 0.2, 0.3, 0.4 measured between 1.75 K and 10 K in an external magnetic field of $\mu_0 H$ = 1 mT in zero-field-cooled (ZFC) mode. The onsets of the transition to superconductivity $T_{c,onset}$ are marked with arrows. The samples $x$ = 0.3 and 0.4 do not display a bulk diamagnetic shielding, and their magnetic susceptibilities are displayed with dashed lines.}
\label{fig:3}
\end{figure}

The DC magnetic susceptibility $\chi(T)$ for temperatures from 1.8 K to 10 K, measured in zero-field cooled (ZFC) mode in an external field of $\mu_0 H$ = 1 mT for the samples $x$ = 0.05, 0.1, 0.15, 0.2, 0.3, and 0.4, are shown in the inset of figure \ref{fig:3}. Almost perfect diamagnetic shielding ($\chi = -1$) is observed for the samples $x$ = 0.05, 0.1, 0.15, and 0.2 at 1.8 K. The samples with $x$ = 0.3 and $x$ = 0.4 undergo a transition to a superconducting state at nearly the same temperature as the optimally doped sample with $x$ = 0.2. However, their diamagnetic shielding fractions are clearly lowered, leading to the conclusion that doping levels beyond $x > $ 0.2 do not display bulk superconductivity anymore (dashed lines). The onset temperatures of the transition to the superconducting states, $T_{c,onset}$, for the bulk superconducting samples are marked with arrows in the corresponding colors.

In figure \ref{fig:3} we summarize the electronic phase diagram of the \ce{Ba_{1-\textit{x}}Rb_{\textit{x}}Ti2Sb2O} solid solution, with the superconducting critical temperature $T_c$ for the parent compound \ce{BaTi2Sb2O} taken from reference \onlinecite{BaTi2Sb2O}. The critical temperatures $T_c$ used for the phase diagram are the intersections of the slopes of the phase transitions and the normal state magnetizations. The observed phase diagram for rubidium substituted \ce{Ba_{1-\textit{x}}Rb_{\textit{x}}Ti2Sb2O} is in good agreement with the earlier reported phase diagrams for sodium, potassium, and bismuth substituted \ce{BaTi2Sb2O}, with the exception that $T_{CDW}$ decreases more rapidly than in the latter compounds.\\
\begin{figure}
\centering
\includegraphics[width=\linewidth]{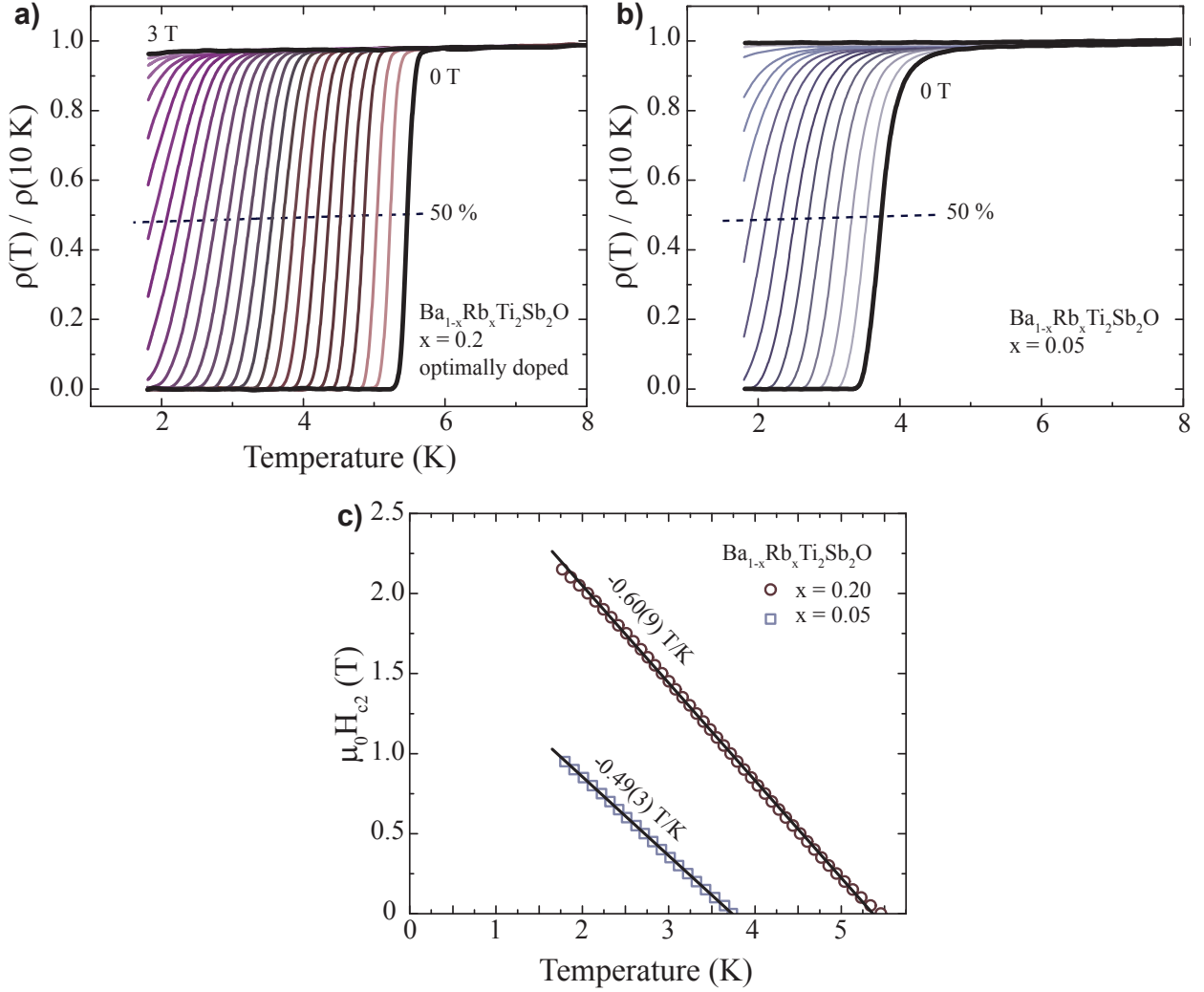}
\caption{(a) and (b) Normalized resistivity $\rho(T)/\rho(10 K)$ for the optimally doped sample $x$ = 0.2 and $x$ = 0.05, respectively. Field-dependent resistivity measurements are shown for magnetic fields between 0 and 3 T, varied by 0.05 T steps, in a temperature range between 1.8 K and 8 K. The dashed lines denote the used 50\% criterion to determine $H_{c2}(T)$, shown in panel (c). The solid lines indicate the extrapolated slopes $\frac{dH_{c2}}{dT}$ used for the WHH approximation (Eq. \ref{eq:WHH}).}
\label{fig:4}
\end{figure}
The field-dependent resistivity measurements for $x$ = 0.05 and optimally doped $x$ = 0.2 are shown in figures \ref{fig:4}a and b for external fields $\mu_0H \leq$ 3 T. As expected, the transition temperature $T_c$ is gradually reduced with increasing magnetic fields. We have defined the corresponding upper-critical fields $H_{c2}$ using a 50\% criterion, i.e., the upper critical field $H_{c2}$(T) is defined by the temperature $T$ at which 50\% of the normal-state
resistivity is suppressed (see dashed lines in figures \ref{fig:4}a and b). The resulting temperature dependences of $\mu_0 H_{c2}$(T) are shown in figure \ref{fig:4}c. The extrapolated slopes are for $x$ = 0.05, $\frac{dH_{c2}}{dT}$ = -0.49(3) T/K, and for the optimally doped sample with $x$ = 0.2, $\frac{dH_{c2}}{dT}$ = -0.60(9) T/K. From these slopes we can evaluate the upper critical fields at zero temperature $\mu_0 H_{c2}$(0) by applying the Werthamer-Helfand-Hohenberg (WHH) approximation,\cite{WHH}
\begin{equation}
H^{WHH}_{c2}(0) = -0.69 \ T_c \ \left(\frac{dH_{c2}}{dT}\right)_{T = T_c}.
\label{eq:WHH}
\end{equation}
The resulting upper critical fields are $\mu_0 H_{c,2}(0) \approx$ 1.3 T for $x$ = 0.05, and $\mu_0 H_{c,2}(0) \approx$ 2.3 T for $x$ = 0.2, respectively. \\
According to Ginzburg-Landau theory, the upper critical field at $T$ = 0 K, $H_{c2}(0)$, can be used to estimate the coherence length $\xi(0)$ at $T$ = 0 K using
\begin{equation}
\mu_0 H_{c2}(0) = \frac{\Phi_0}{2 \pi \ \xi(0)^2},
\label{eq:GL}
\end{equation}
with $\Phi_0 = h/(2e) \approx 2.0678 \cdot 10^{-15}$ Wb being the magnetic flux quantum. We obtain coherence lengths $\xi(0)$ = 160 $\mathrm{\AA}$ for $x$ = 0.05, and $\xi(0)$ = 120 $\mathrm{\AA}$ for the optimally doped sample with $x$ = 0.2. Recently, Gooch et al. \cite{Gooch13} derived from field dependent specific-heat measurements for sodium substituted \ce{Ba_{0.85}Na_{0.15}Ti2Sb2O} with a $T_c$ = 4.2 K an upper critical field $H_{c2}(0)$ = 1.7 T, and a corresponding coherence length $\xi(0)$ = 140 $\mathrm{\AA}$. Moreover, an upper-critical field $H_{c2}(0)$ = 0.08 T and a corresponding coherence length $\xi(0)$ = 640 $\mathrm{\AA}$ were reported for the parent compound \ce{BaTi2Sb2O}. The coherence lengths $\xi(0)$ and the upper-critical fields $H_{c2}(0)$ that we estimated here, for the rubidium doped \ce{BaTi2Sb2O}, are in good agreement with these earlier reported values. The layered structure of these materials implies, as it is the case for the cuprates, that the physical properties are highly anisotropic. Therefore, these numbers for $H_{c2}$ and the coherence lengths $\xi$ obtained from polycrystalline samples have to be taken as values appropriately averaged over all crystal directions.
\\
\begin{figure}
\centering
\includegraphics[width=0.8\linewidth]{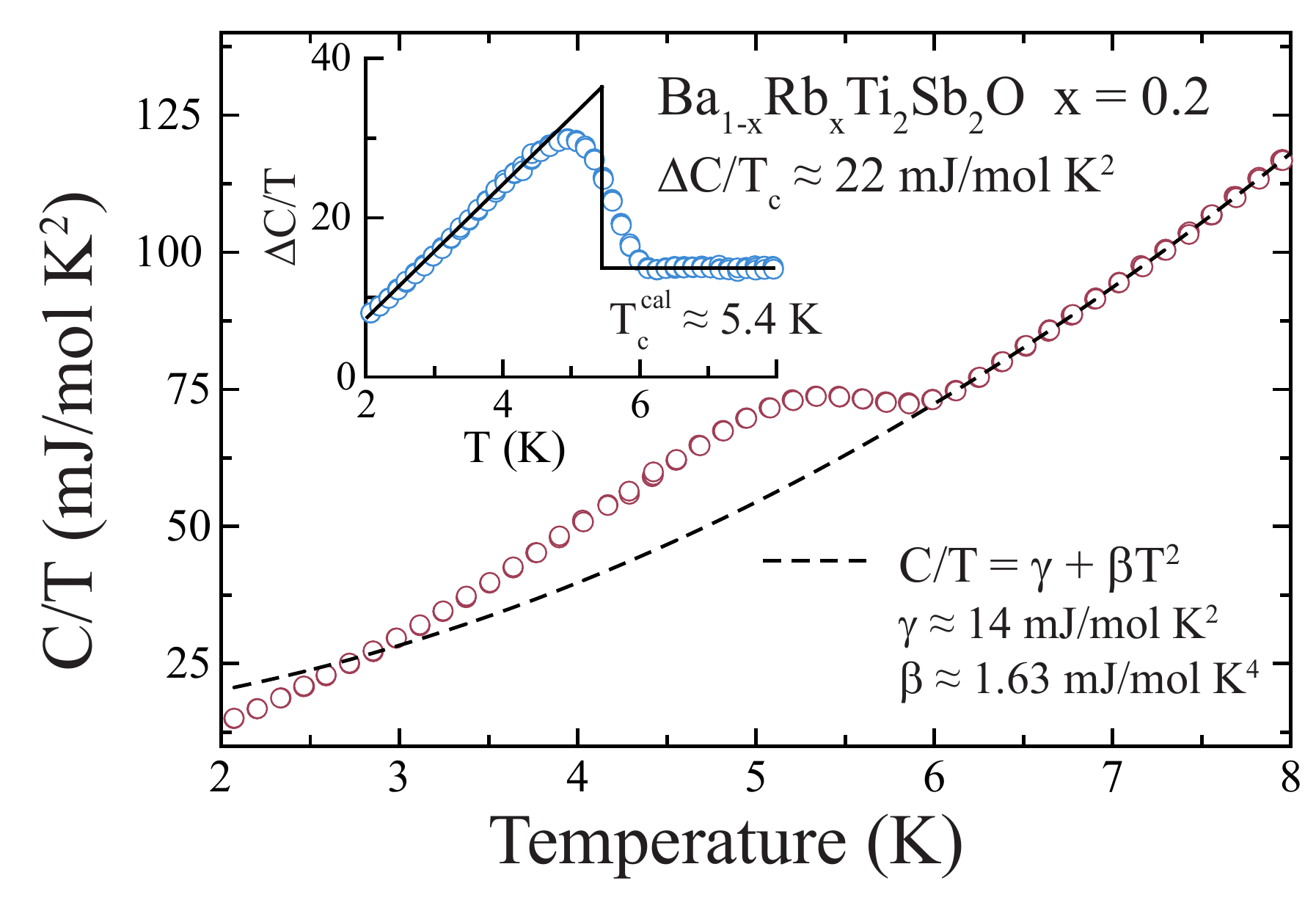}
\caption{Reduced specific heat $C/T$ vs. temperature $T$ of the optimally doped sample ($x$ = 0.2), together with an inset showing the same data after subtraction of the normal-state contribution (dashed line in the main panel, see text). The solid lines in the inset represent an entropy-conserving construction to obtain the discontinuity and $\Delta C/T_c$ and $T^{cal}_c$.}
\label{fig:5}
\end{figure}

In figure \ref{fig:5} we show the specific heat of the optimally doped sample ($x$ = 0.2), in a $C/T$ vs. $T$ representation. The normal-state contribution has been fitted to the data between $T$ = 6 K and 13 K according to the standard expression 
\begin{equation}
\frac{C(T)}{T} = \gamma + \beta T^2 
\end{equation}
with the Sommerfeld constant $\gamma$ and $\beta = 12 \pi^4 n R / 5 \Theta^3_D$, where $n$ = 6 is the number of atoms per formula unit, R = 8.13 J/mol K is the gas constant, and $\Theta_D$ the Debye temperature (dashed line). The obtained value $\gamma \approx $ 14 mJ/mol K$^2$ is in line with corresponding values reported for \ce{BaTi2Sb2O}\cite{BaTi2Sb2O} and \ce{Ba_{0.85}Na_{0.15}Ti2Sb2O} \cite{Gooch13}, with $\gamma$ values ranging between 10 and 15 mJ/mol K$^2$. In the inset of figure \ref{fig:5} we show the measured $\Delta C/T$ data, with this normal-state contribution subtracted, together with an entropy-conserving construction to obtain the calorimetrically determined critical temperature $T^{cal}_c$ and the discontinuity $\Delta C/T_c$ at $T_c$. With $\Delta C/T_c \approx$ 22 mJ/mol K$^2$ at $T^{cal}_c$ = 5.4 K we obtain a ratio $\Delta C / \gamma T_c \approx$ 1.6, which only slightly exceeds the standard weak-coupling BCS value, $\Delta C / \gamma T_c =$ 1.43, again in qualitative agreement with the reported data for \ce{Ba_{0.85}Na_{0.15}Ti2Sb2O} \cite{Gooch13}. A $\beta \approx$ 1.63 mJ/mol K$^4$ corresponds to $\Theta_D \approx$ 193 K, which is somewhat smaller than $\Theta_D \approx$ 210-239 K as communicated for \ce{BaTi2Sb2O} and \ce{Ba_{0.85}Na_{0.15}Ti2Sb2O}.\cite{BaTi2Sb2O,Gooch13}\\
\begin{figure}
\centering
\includegraphics[width=\linewidth]{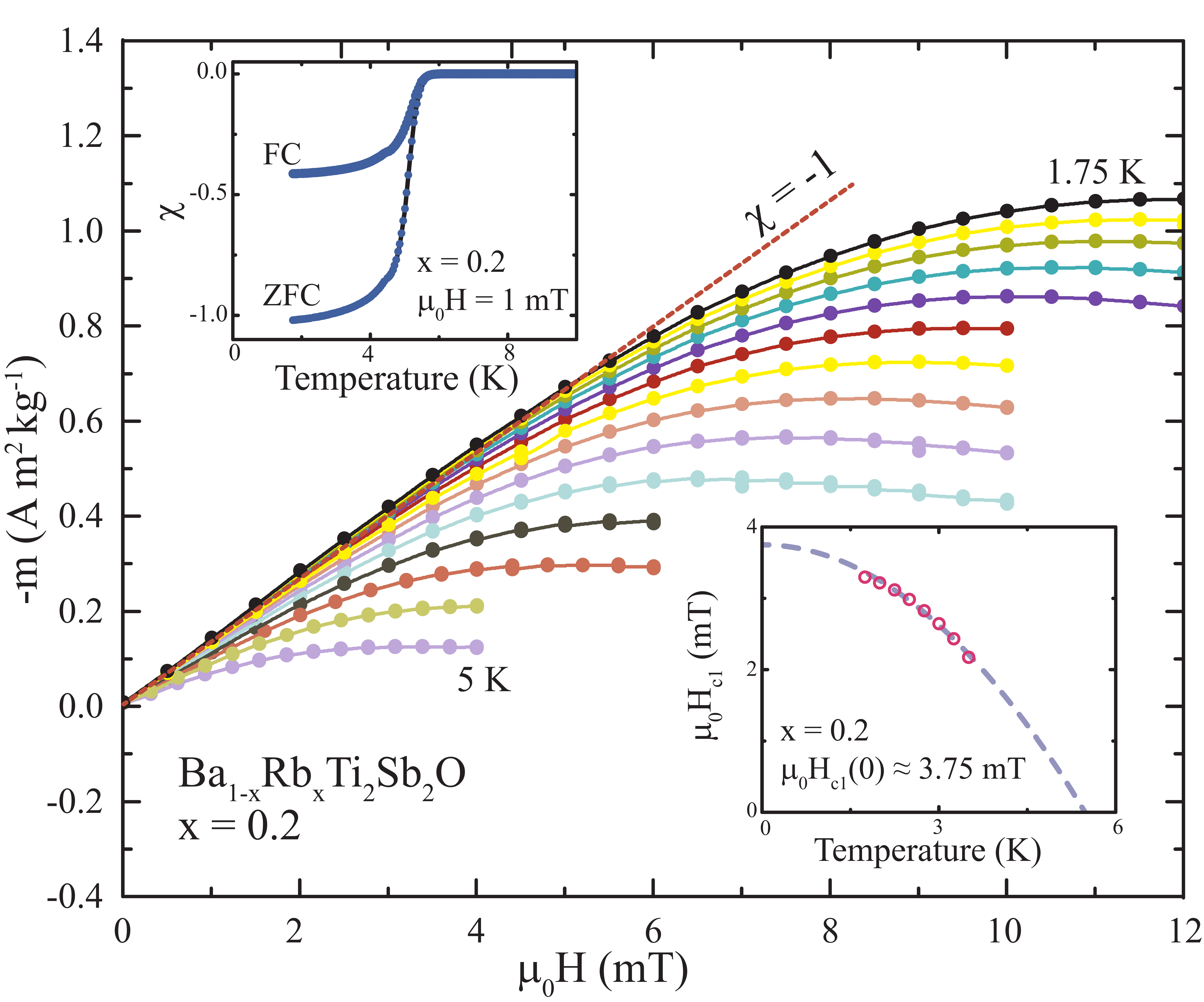}
\caption{ZFC field dependence of the magnetization $m(H)$ of the optimally doped sample with $x$ = 0.2, for temperatures between 1.75 K and 5 K (in 0.25 K steps), in magnetic fields $\mu_0 H$ between 0 and 15 mT. The dashed line shows the ideal diamagnetic shielding. Upper inset: ZFC and FC magnetic susceptibility of the sample with $x$ = 0.2. Lower inset: The temperature dependence of the lower critical field $H_{c1}$ of the sample with $x$ = 0.2 (see text). The dashed line is a fit to equation \ref{eq:hc1} with $T_c$ fixed to 5.4 K.}
\label{fig:6}
\end{figure}
In the upper inset of figure \ref{fig:6} we show the zero-field cooled (ZFC) and the field-cooled magnetic susceptibilities for the optimally doped sample $x$ = 0.2. The large screening efficiency, as well as a large Meissner shielding of the order of $\approx$ 40 \% are further indicators for the bulk nature of superconductivity in this compound. The ZFC field dependence of the magnetization $m(H)$ for temperatures between 1.75 K and 5 K (in 0.25 K steps) in magnetic fields $\mu_0 H$ between 0 and 15 mT are shown in the main panel of figure \ref{fig:6}. For comparison, the ideal linear behavior expected in the Meissner state ($\chi $ = -1) is also shown (dashed line). It is very difficult to extract precise values for the lower critical field $H_{c1}$ from $m(H)$ measurements, especially for polycrystalline samples. A criterion that is often used in the literature, namely the identification of $H_{c1}$ as the magnetic field where $m(H)$ first deviates from linearity, does not rely on a sharp feature in the experimental data. A somewhat better criterion was given in reference \onlinecite{hc1} based on Bean's critical-state model,\cite{bean} although the definition of a linear $m(H)$ regime used in that approach also leaves some room of ambiguity. We nevertheless applied the procedure described in reference \onlinecite{hc1} to selected $m(H)$ data for which we could identify a clearly linear $m(H)$ in the limit $H\rightarrow 0$ (i.e., for our low-temperature data), and the resulting $H_{c1}$ values are plotted in the lower inset of figure \ref{fig:6}. A reasonable estimate for $\mu_0 H_{c1}(0)$ can then be obtained by using an empirical formula,\cite{Brandt}
\begin{equation}
H_{c1}(T) = H_{c1}(0) [1-(T/T_c)^2].
\label{eq:hc1}
\end{equation}
With this approximation and fixing $T_c$ = 5.4 K (see dashed line in lower inset of figure \ref{fig:6}) we obtain $\mu_0 H_{c1}(0) \approx$ 3.8 mT. Depending on the criterion used for defining $H_{c1}$, these numbers may vary by a factor of unity (taking, for example, the maximum of $-m(H)$ as a criterion for $H_{c1}$, yields corresponding values that are larger by a factor of approximately 4). Moreover, all these numbers represent again values that are averaged over all crystal directions. We can nevertheless state that these materials must be strongly type-II superconductors with a Ginzburg-Landau parameter of the order of $\kappa = \lambda/\xi \approx$ 35 for $x$ = 0.2, as estimated from the relations \cite{Brandt}
\begin{equation}
\mu_0 H_{c1} = \frac{\phi_0}{4 \pi \lambda^2} ln(\kappa + 0.5),  
\end{equation}
\begin{equation}
\mathrm{and} \quad \frac{H_{c1}}{H_{c2}} = \frac{ln(\kappa)+ \frac{1}{2}}{\kappa^2}  
\end{equation}
with a $\lambda \approx 4200 \mathrm{\AA}$.
\section{Conclusion}
We have described the successful synthesis of the \ce{Ba_{1-\textit{x}}Rb_{\textit{x}}Ti2Sb2O} $x \leq 0.4$ solid solution, and presented data on their basic physical properties. X-ray diffraction data show that the compounds are single phase with a crystal structure isopointal to the \ce{CeCr2Si2C}-type (\textit{P}4/\textit{mmm}) structure. Our temperature dependent resistivity measurements reveal a continuous and drastic decrease of the CDW ordering transition temperature, by replacing barium by rubidium, with the CDW transition fully suppressed for $x \geq 0.2$. At this doping level, superconductivity reaches its maximum critical temperature $T_c$ = 5.4 K (in specific-heat measurements). Both a well developed discontinuity in the specific heat at $T_c$, and a large Meissner-shielding fraction indicate the bulk nature of superconductivity. Larger rubidium contents than the optimum value $x = 0.2$ lead to a decrease of the diamagnetic shielding fraction, with a $T_c$ remaining essentially unchanged. From our data we obtain estimates for the lower and the upper critical fields $H_{c1}(0) \approx$ 3.8 mT and $\mu_0 H_{c,2}(0) \approx$ 2.3 T for the optimally doped sample with $x$ = 0.2, indicating that these compounds are strongly-type II superconductors.
Our results support the scenario that hole doping by the incorporation of Na$^+$, K$^+$, and also Rb$^+$ is of great importance for the suppression of CDW ordering and the occurrence of superconductivity in these materials.
\section*{Acknowledgments}
FvR acknowledges a scholarship from Forschungskredit UZH, grant no. 57161402. The authors would like to thank Michael W\"orle for helpful discussion.

 \end{document}